\documentclass[12pt]{article}
\usepackage{a4,amsmath}
\begin{document}
\begin{center}
\begin{Large}
{\bf{Lorentz symmetry violation without violation of relativistic symmetry}} 
\end{Large}   
\vskip 5mm
\begin{large}
      G.Yu. Bogoslovsky \\
\end{large}
\vskip 3mm
{\footnotesize{{\it{Skobeltsyn Institute of Nuclear Physics, Moscow State University, 119992 Moscow, Russia\\
E-mail address:}} bogoslov@theory.sinp.msu.ru}}\\
\end{center}
\vskip 7mm
\hrule
\vskip 3mm
\begin{small}
\noindent
{\bf{Abstract}}
\vskip 2mm
  Within the framework of the Finslerian approach to the problem of
violation of Lorentz symmetry, consideration is given to a flat
axially symmetric Finslerian space of events, which is the
generalization of Minkowski space. Such an event space arises from the
spontaneous breaking of initial gauge symmetry and from the formation
of anisotropic fermion-antifermion condensate. It is shown that the
appearance of an anisotropic condensate breaks Lorentz symmetry;
relativistic symmetry, realized by means of the 3-parameter group of
generalized Lorentz boosts, remains valid here nevertheless. We have obtained the
bispinor representation of the group of generalized Lorentz boosts,
which makes it possible to construct the Lagrangian for an interaction
of fundamental fields with anisotropic condensate.\\
\end{small} 
\vskip 3mm
{\footnotesize{   
\noindent
{\sl{PACS\,:}}\, 03.30.+p; 11.30.Cp; 11.30.Qc; 11.15.Ex; 12.10.Dm; 02.20.Sv; 02.40.-k
\vskip 1mm
\noindent
\phantom{{\it{Keywords}}\,: }Finslerian space-time}}                 
\vskip 3mm
\hrule
\vskip 2cm

\noindent
{\large{\bf 1\,. Introduction}}

\bigskip

\noindent
It is known that the Einstein relativity principle is most fully
reflected in the requirement of local Lorentz invariance of the laws
of nature. Compared with the other physical symmetries, Lorentz
space-time symmetry occupies a special place: since its discovery a
hundred years ago it has determined the development of the
theory of fundamental interactions. At present, however, owing to
progress made in the construction of unified gauge theories, there
have been reasons to consider Lorentz symmetry not as a strict but
only as an approximate symmetry of nature. 

Subsequent to the paper [1], the following point of view is becoming
more and more popular: in this view even in a theory which has Lorentz
invariance at the most fundamental level, this symmetry can be
spontaneously broken if some (for example, vector) field acquires a
vacuum expectation value which breaks the initial Lorentz
symmetry. Certainly the question here concerns symmetry violation with
respect to active Lorentz transformations of fundamental fields
against the background of a fixed (in this case, vector) condensate. As
for passive Lorentz transformations, under which the condensate is
transformed as a Lorentz vector, 
the corresponding Lorentz covariance remains valid. 

It is a noteworthy fact that the usual Standard Model of strong, 
weak and electromagnetic interactions does not possess
the dynamics necessary to  cause spontaneous breaking of Lorentz
symmetry. In other words, the standard Higgs mechanism, which breaks
the local gauge symmetry and gives rise to the scalar Higgs condensate,
does not affect the initial Lorentz symmetry of the theory. However,
the fact that in more complicated theories, in particular string
theories, spontaneous breaking of Lorentz symmetry may well occur has given
impetus to the construction of a phenomenological theory [2] referred
to as the Standard Model Extension (SME). The Lagrangian of this
theory includes all passive Lorentz scalars formed by combining
standard-model fields with coupling coefficients having Lorentz
indices. As for all dominant gravitational couplings in the SME
action, such a problem has been studied in [3]. As a result the SME
has made it possible to describe many possible effects caused by
violation of active Lorentz invariance and to classify them as effects
of Planck-scale physics, strongly suppressed at attainable energy scales. 

In spite of the absence of direct evidence in favour of Lorentz
symmetry violation, interest in the given fundamental problem is still
increasing [4]. In this connection we also point out the publications [5,6]
in which, along with experimental investigations inspired, in
particular, by the SME, other theoretical approaches to Lorentz
violation are also discussed. One of such approaches was proposed
relatively long ago [7]; it is based on the more general,
Finslerian geometrical model of space-time [8,9] rather than on the
Riemannian one. Although from the phenomenological point of view the
Finslerian approach is still in need of further development,
potentially it is more appealing since it admits violation of 
Lorentz invariance with no breaking of relativistic invariance. 

The key moment in the formation of the Finslerian approach was the
fact that within the framework of the principle of correspondence with
Minkowski metric it became possible to find two types of Finslerian
metrics, which possess relativistic symmetry, i.e.\ symmetry
corresponding to boosts. The first of the above-mentioned metrics
describes a flat space of events with axial symmetry, i.e.\ with
partially broken $3D$  isotropy [7] while the second 
one [10] exhibits an entirely broken $3D$ isotropy. 

The appearance of $3D$ anisotropy of the event space with the preservation
of its re\-la\-tivistic symmetry indicates that the spontaneous
breaking of the initial gauge symmetry may be accompanied by a 
corresponding phase transition in the geometrical structure of
space-time. In other words, spontaneous breaking of gauge symmetry may
lead to a dynamic rearrangement of vacuum which results in the
formation of a relativistically invariant anisotropic
fermion-antifermion condensate, i.e.\ of a constant classical
nonscalar field. This constant field physically manifests itself as a
relativistically invariant anisotropic medium filling space-time. Such
a medium, leaving space-time flat, gives rise to its anisotropy, that
is, instead of Minkowski space there appears a relativistically
invariant Finslerian event space. Since the relativistic symmetry of
the Finslerian space is realized with the aid of the so-called
generalized Lorentz transfomations, being different from the
usual Lorentz boosts, this then is why one can speak of violation of 
Lorentz symmetry, too. However the remaining relativistic
symmetry, now represented by a 3-parameter noncompact group of the
generalized Lorentz transformations, still plays the important
constructive role. In particular, it makes possible to take
expli\-citly into account the influence of the anisotropic
fermion-antifermion condensate on the dynamics of fundamental fields 
after spontaneous breaking of the initial gauge symmetry. 

In view of the foregoing we call attention to the fact that,
irrespective of the problem of violation of Lorentz symmetry, in the
literature consideration has already been given to the mechanism of
the dynamical breaking of the initial gauge symmetry which is alternative
to the standard one; instead of the elementary Higgs condensate there
appears a scalar fermion-antifermion condensate [11]. As for
fermion-antifermion condensate, which would break $3D$ isotropy of
event space in a relativistically invariant way, it should be noted
that in terms of quantum theory such a problem has not yet been
considered. However, within the framework of classical theory [12],
relativistically invariant fermion-antifermion 
condensate, which leads at least to partial breaking of $3D$ isotropy, actually arises. 

In the present work we confine ourselves to the case of flat
space-time with partially broken $3D$ isotropy and again consider the
group of its relativistic symmetry. Thereafter we shall construct the
bispinor representation of the corresponding group. 

In comparison with the 6-parameter homogeneous Lorentz group of
Minkowski space, the homogeneous group of isometries of the flat
Finslerian space with partially broken $3D$ isotropy is a 4-parameter group. 
Apart from 3-parameter boosts (generalized Lorentz transformations) it 
includes only axial symmetry transformations, i.e.\ the 1-parameter
group of rotations about the preferred direction in $3D$ space; in
this case this direction is determined by a spontaneously arising
axially symmetric fermion-antifermion condensate. 

It will be demonstrated below that the given 4-parameter group of
Finslerian isometries is locally isomorphic to the corresponding
4-parameter subgroup of the Lorentz group. Since this fact is fundamental
for the construction of the bispinor representation of the group of Finslerian
isometries, we primarily consider the above-mentioned 4-parameter
subgroup of the Lorentz group. 

\bigskip
\bigskip

\noindent
{\large{\bf 2\,. The 4-parameter subgroup of Lorentz group\\
    \phantom{and}and its 3-parameter noncompact subgroup}}\\

\bigskip

\noindent
In [13], in terms of Lie algebras all continuous subgroups of Lorentz
group were classified. It turned out that the Lorentz group contains
not a single 5-parameter subgroup and has a single (up to isomorphism)
4-parameter subgroup. This subgroup includes independent rotations
about an arbitrarily selected axis, the direction of which will be
denoted using a unit vector $\boldsymbol\nu$, and a 3-parameter group
consisting of noncompact transformations only. Physically such
noncompact transformations are realized as follows. First choose as
$\boldsymbol\nu$ a direction towards a preselected star and then
perform an arbitrary Lorentz boost by complementing it with such a
turn of the spatial axes that in a new 
reference frame the direction towards the star remains unchanged. 

The set of the transformations described, while linking the inertial
reference frames, actually constitutes a 3-parameter noncompact group
(in contrast to the usual Lorentz boosts). Let us write the
corresponding 3-parameter transformations in the infinitesimal 
form 
\begin{eqnarray}\label{1}
dx^0 &=& - \boldsymbol{n} \boldsymbol{x} d \alpha, \nonumber\\
d\boldsymbol{x} &=& (- \boldsymbol{n} x^0 - [\boldsymbol{x}[\boldsymbol{\nu}\boldsymbol{n}]])d\alpha
\end{eqnarray}
where $d\alpha$  is a rapidity, the unit vector $\boldsymbol{n}$ indicates a
direction of the infinitesimal boost, so that $d\boldsymbol{v} =
\boldsymbol{n}d\alpha$, and the meaning of $\boldsymbol{\nu}$ has been
explained. Integration of equations (1) leads to final transformations
which, at any fixed $\boldsymbol\nu$, belong to Lorentz group 
and themselves form a 3-parameter noncompact group with parameters
${\boldsymbol n}\,,\alpha\,$\,:
\begin{equation} 
\label{2}
x^{'i} =\Lambda^i_k(\boldsymbol{\nu};\boldsymbol{n},\alpha )\,x^k\,, 
\end{equation}
where
\begin{eqnarray*}
\Lambda^0_0&=&1+\frac{\cosh{\boldsymbol{\nu n}\alpha}-1}{(\boldsymbol{\nu n})^2}\,,\\
\Lambda^0_{\beta}&=&\frac{1-e^{-\boldsymbol{\nu n}\alpha}}{\boldsymbol{\nu n}}\,n_{\beta}+\frac{\cosh{\boldsymbol{\nu n}\alpha}-1}{(\boldsymbol{\nu n})^2}\,\nu_{\beta}\,,\\
\Lambda^{\rho}_0&=&\frac{1-e^{\boldsymbol{\nu n}\alpha}}{\boldsymbol{\nu n}}\,n^{\rho}+\frac{\cosh{\boldsymbol{\nu n}\alpha}-1}{(\boldsymbol{\nu n})^2}\,\nu^{\rho}\,,\\
\Lambda^{\rho}_{\beta}&=&\delta^{\rho}_{\beta}+\frac{1-e^{\boldsymbol{\nu n}\alpha}}{\boldsymbol{\nu n}}\,n^{\rho}\nu_{\beta}+\nu^{\,\rho}\Lambda^0_{\beta}\,.
\end{eqnarray*} 
Hereafter the latin indices take on values of 0, 1, 2, 3 while the greek ones, values of 1, 2, 3. Note also that the 
$n^{\,\beta}$ and ${\nu}^{\,\beta}$ denote the Cartesian components of unit vectors  ${\boldsymbol n} $ and    
$\boldsymbol\nu\,$, in which case $n_{\,\beta}=-n^{\,\beta}\,,\ {\nu}_{\,\beta}=-{\nu}^{\,\beta}\,$. The transformations inverse to (2) appear as 
\begin{equation} 
\label{3}
x^{i} ={\Lambda^{-1}}^i_k(\boldsymbol{\nu};\boldsymbol{n},\alpha)\,x^{'k}\,, 
\end{equation}
where
\begin{equation} 
\label{4}
{\Lambda^{-1}}^i_k(\boldsymbol{\nu};\boldsymbol{n},\alpha )=\Lambda^i_k
(\boldsymbol{\nu};\boldsymbol{n},-\alpha)\,.
\end{equation}

Consider two arbitrary elements of the group (2). Let the first element 
$g_1$ be characterized by the pa\-ra\-me\-ters ${\boldsymbol n}_1\,,  \alpha_1\,$, and the second one, $g_2\,,$ by the pa\-ra\-meters ${\boldsymbol n}_2\,, \alpha_2\,$. Then to the element   $g=g_2g_1$ there will correspond the parameters ${\boldsymbol n}\,, 
\alpha\,$, which are functionally dependent on the ${\boldsymbol n}_1\,, \alpha_1$ and  ${\boldsymbol n}_2\,, \alpha_2\,$, i.e.\ 
$\Lambda^i_k(\boldsymbol{\nu};\boldsymbol{n},\alpha)=
\Lambda^i_j(\boldsymbol{\nu};{\boldsymbol n}_2,{\alpha}_2)
\Lambda^j_k(\boldsymbol{\nu};{\boldsymbol n}_1,{\alpha}_1)\,.$ Using the explicit form of the matrix elements $\Lambda_k^i(\boldsymbol {\nu};{\boldsymbol n},\boldsymbol\alpha)$ and making the corresponding calculations we arrive at the following relations: 
\begin{equation} 
\label{5}
{\boldsymbol n}\,\alpha = \frac{\boldsymbol\nu ({\boldsymbol n}_1{\alpha}_1+{\boldsymbol n}_2{\alpha}_2)}{1-e^{\boldsymbol\nu ({\boldsymbol n}_1{\alpha}_1+{\boldsymbol n}_2{\alpha}_2)}}\left[\frac{1-e^{\boldsymbol\nu{\boldsymbol n}_1{\alpha}_1}}{\boldsymbol\nu{\boldsymbol n}_1}\,{\boldsymbol n}_1 + \frac{e^{\boldsymbol\nu{\boldsymbol n}_1{\alpha}_1}(1-e^{\boldsymbol\nu{\boldsymbol n}_2{\alpha}_2})}{\boldsymbol\nu{\boldsymbol n}_2}\,{\boldsymbol n}_2\right]\,,
\end{equation}

\begin{equation}
\label{6}
{\boldsymbol n}^2 = 1\,.
\end{equation}
These relations essentially represent the law of group composition for the 3-parameter noncompact subgroup (2) of Lorentz group. 

Since the group (2) links the coordinates of events in the initial and primed inertial reference frames, from the physical standpoint it is more natural to use as group parameters three velocity components, $\boldsymbol v$, of the primed reference frame rather than the $\boldsymbol{n}, \alpha$\,. In order to express the $\boldsymbol v$ in terms of the $\boldsymbol{n}, \alpha$ it is sufficient to consider motion in the initial frame of the origin of the primed frame, i.e.\ to write 
$x^{\,\beta}={\Lambda^{-1}}^{\beta}_0\,x^{'\,0}$ and 
$x^{\,0}={\Lambda^{-1}}^0_0\,x^{'\,0}\,$. Then $v^{\,\beta}=x^{\,\beta}/x^{\,0}={\Lambda^{-1}}^{\beta}_0/
{\Lambda^{-1}}^0_0\,\,$. Using now eqs.\ (2)--(4), we get as a result 
\begin{equation} 
\label{7}
\boldsymbol v=\left[\frac{1-e^{-\boldsymbol{\nu n}\alpha}}{\boldsymbol{\nu n}}\,
\boldsymbol n +\frac{\cosh{\boldsymbol{\nu n}\alpha}-1}{(\boldsymbol{\nu n})^2}\,
\boldsymbol\nu \right]\biggl/\left[1+\frac{\cosh{\boldsymbol{\nu n}\alpha}-1}
{(\boldsymbol{\nu n})^2}\right]\,.
\end{equation}
Hereafter we put  $c=1$. Since ${\boldsymbol n}^2 = {\boldsymbol\nu}^2 = 1\,$, eq.\ (7) yields the inverse relations:
\begin{equation} 
\label{8}
\boldsymbol n =\frac{\boldsymbol v}{\sqrt{2(1-\boldsymbol{v\nu })(1-\sqrt{1-
{\boldsymbol v}^2})}}\, - \,\sqrt{\frac{1-\sqrt{1-{\boldsymbol v}^2}}
{2(1-\boldsymbol{v\nu })}}\,\,\,\boldsymbol{\nu}\,,
\end{equation}

\begin{equation} 
\label{9}
\alpha =\frac{\sqrt{2(1-\boldsymbol{v\nu })(1-\sqrt{1-{\boldsymbol v}^2})}}
{\sqrt{1-{\boldsymbol v}^2}+\boldsymbol{v\nu }-1}\, 
\ln\left(\frac{\sqrt{1-{\boldsymbol v}^2}}{1-\boldsymbol{v\nu }}\right)\,.
\end{equation}
In terms of $\boldsymbol v$ the law of group composition (5),(6) takes the form 
\begin{equation} 
\label{10}
\boldsymbol v =\frac{({\boldsymbol v}_1(1\!-\!
{\boldsymbol v}_2\boldsymbol\nu )\!+\!{\boldsymbol v}_2\sqrt{1\!-\!{\boldsymbol v}_1^2}
)(1\!-\!{\boldsymbol v}_1
\boldsymbol\nu )\!+\!\boldsymbol\nu ({\boldsymbol v}_1
{\boldsymbol v}_2\!+\!\boldsymbol\nu{\boldsymbol v}_2 (\sqrt{1\!-\!
{\boldsymbol v}_1^2}-\!1))\sqrt{1\!-\!{\boldsymbol v}_1^2}}{1-{\boldsymbol v}_1
\boldsymbol\nu +{\boldsymbol v}_1{\boldsymbol v}_2\sqrt{1-{\boldsymbol v}_1^2}+
\boldsymbol\nu{\boldsymbol v}_2(1-{\boldsymbol v}_1\boldsymbol\nu +\sqrt{1-
{\boldsymbol v}_1^2})(\sqrt{1-{\boldsymbol v}_1^2}-1)}\,, 
\end{equation}
whereby $\Lambda^i_k(\boldsymbol{\nu};\,\boldsymbol{v})=
\Lambda^i_j(\boldsymbol{\nu};\,{\boldsymbol v}_2)
\Lambda^j_k(\boldsymbol{\nu};\,{\boldsymbol v}_1)\,$. It is thus clear that eq.\ (10) represents Einstein's law of addition of velocities  ${\boldsymbol v}_1$ and  ${\boldsymbol v}_2\,$. In comparison with its usual form one should remember, however, that after the transformation 
$\Lambda^j_k(\boldsymbol{\nu};\,{\boldsymbol v}_1)$ the spatial axes, in which the ${\boldsymbol v}_2\,$ is prescribed, appear now not to be parallel to the initial axes but turned so that the vector $\boldsymbol\nu$ relative to them maintains its initial orientation. It is just therefore, irrespective of the direction of 
${\boldsymbol v}_1\,$, eq.\ (10) yields $\boldsymbol v = \boldsymbol\nu\,$ if  ${\boldsymbol v}_2 = {\boldsymbol\nu\,}$.

\bigskip
\bigskip

\noindent
{\large{\bf 3\,. Matrices of the finite bispinor transformations\\ \phantom{rep}representing a 3-parameter group\\ \phantom{of t}of the generalized Lorentz transformations}}

\bigskip

\noindent
The 3-parameter group of the generalized Lorentz transformations, similarly to the subgroup (2) of Lorentz group, consists of noncompact transformations only. In the infinitesimal form the transformations belonging to it appear as 
\begin{eqnarray}\label{11}
dx^0 &=& (-r(\boldsymbol{\nu}\boldsymbol{n})x^0 - \boldsymbol{n} \boldsymbol{x}) d \alpha\,, \nonumber\\
d\boldsymbol{x} &=& (-r(\boldsymbol{\nu}\boldsymbol{n})\boldsymbol{x} - \boldsymbol{n} x^0 - [\boldsymbol{x}[\boldsymbol{\nu}\boldsymbol{n}]])d\alpha\,.
\end{eqnarray}
Here, as in the infinitesimal transformations (1) of the group (2), the 
${\boldsymbol n}$ and  $\alpha$ are group parameters while the
$\boldsymbol\nu$ is a fixed unit vector. And the difference between
(11) and (1) consists in the appearance of an additional generatior of
the scale transformations, which is proportional to a new fixed
dimensionless parameter $r\,$. Since the scale transformations commute
with the Lorentz 
boosts and $3D$ rotations, the result of integration of eqs.\ (11) is a priori clear:
\begin{equation} 
\label{12}
x^{'i} =D(r,\boldsymbol{\nu};\,\boldsymbol{n},\alpha)\Lambda^i_k(\boldsymbol{\nu};\boldsymbol{n},\alpha)\,x^k\,, 
\end{equation}
where $D(r,\boldsymbol{\nu};\,\boldsymbol{n},\alpha)=e^{-r\,{\boldsymbol\nu}{\boldsymbol n}\,\alpha}I\,$, whereby  $I$ is a unit matrix while $\Lambda^i_k(\boldsymbol{\nu};\boldsymbol{n},\alpha )$ are matrices which make up the group (2)\,. As for the law of group composition for the group (12), it is given by eqs.\ (5) and (6) obtained for the group (2). We note incidentally that eq.\ (5) yields the relation 
$\boldsymbol\nu{\boldsymbol n}\,\alpha =
\boldsymbol\nu{\boldsymbol n}_1{\alpha}_1 + 
\boldsymbol\nu{\boldsymbol n}_2{\alpha}_2\,$. Thus the group of the
generalized Lorentz transformations (12), on the one hand, is locally
isomorphic to the corresponding 3-parameter subgroup (2) of Lorentz
group and, on the other hand, it is a homogeneous 3-parameter
noncompact subgroup of the similitude group [14]. Since (according to
(12)) in passing to the primed inertial frame the time $(x^0)$ and
space $(\boldsymbol x)$ event coordinates are subjected to identical
additional dilatations $D$\,, then the velocity $\boldsymbol v$ of the
primed frame is related to the group parameters ${\boldsymbol n}\,,
\alpha$ by the same eqs.\ (7)--(9) as in the case $r=0$ where the
group (12) coincides with the subgroup (2) of Lorentz group. For the
same reason the transformations (12) retain valid Einstein's law of
addition of 3-velocities, written as (10). As for the
reparametrization of the group (12), then, for example, the matrix
$D$, involved in (12), takes the following form in 
terms of $\boldsymbol v$:
\begin{equation}
\label{13}
D(r,\boldsymbol\nu;\,\boldsymbol v )=\left(\frac{1-\boldsymbol v\boldsymbol\nu}
{\sqrt{1-\boldsymbol v^{\,2}}}
\right)^r I\,.
\end{equation}

Unlike the transformations (2), 3-parameter group of the generalized Lorentz boosts (12) conformly modifies Minkowski metric but leaves invariant the metric 
\begin{equation}\label{14}
ds^2=\left[\frac{(dx_0-\boldsymbol\nu d\boldsymbol x)^2}{dx_0^2-d\boldsymbol x^{\,2}}\right]^r
(dx_0^2-d\boldsymbol x^{\,2})\,.
\end{equation}
The given Finslerian metric generalizes Minkowski metric and describes the relativistically invariant flat space of events with partially broken $3D$ isotropy. The inhomogeneous isometry group of space (14) is 8-parameter: four parameters correspond to space-time translations, one parameter, to rotations about the physically preferred direction $\boldsymbol\nu$, and three parameters, to the generalized Lorentz boosts. 

Now turn to the construction of bispinor representation  of the group of the ge\-ne\-ra\-lized Lorentz boosts (12). Since the group (12) is locally isomorphic to the 3-parameter subgroup (2) of Lorentz group, its bispinor representation must also be locally isomorphic to the bispinor representation of the subgroup (2). This signifies that the transformations $x'^i = D \Lambda_k^i x^k$ of the event coordinates should be accompanied by the following transformations of a bispinor field: 
$\Psi'(x') = D^dS\Psi(x)$, ${\overline{\Psi}\,}'(x') = \overline{\Psi}(x)D^dS^{-1}$. Here the group of matrices $S$, operating on the bispinor indices, represents the subgroup (2) of the Lorentz matrices $\Lambda_k^i$ while  $D^d$ denotes the corresponding additional scale transformations of bispinors, in which case the unit matrix, involved in the definition (13), operates on the bispinor indices in this context. Since $d^4x' = |D\Lambda_k^i|d^4x = D^4d^4x$ and matrices $S$ satisfy the standard condition $S^{-1}\gamma^n S = \Lambda_m^n\gamma^m$, then, proceeding from the generalized Lorentz invariance of action for a free massless field $\Psi$, it is easy to show that $d = -3/2$. As a result the bispinor representation of the group of generalized Lorentz boosts (12) is realized by the transformations
\begin{equation}
\label{15}
\Psi'= D^{-3/2}S\Psi\,, \qquad {\overline{\Psi}\,}'=\overline{\Psi}D^{-3/2}S^{-1}
\end{equation}
and it remains to find a 3-parameter group of the matrices $S = S(\boldsymbol\nu;\boldsymbol n,\alpha)\,$. For this purpose, using the 4-vectors $\nu^l=(1,\, \boldsymbol{\nu})\,;\, \nu_l=(1,\, -\boldsymbol{\nu})\,;\, n^l=(0,\, \boldsymbol{n})\,;\, n_l=(0,\, -\boldsymbol{n})\,,$ first rewrite the infinitesimal transformations (1) in the form $dx^i = {\omega^i}_k x^k$, where ${\omega^i}_k =(\nu^in_k - n^i \nu_k) d \alpha\,,$ in which case $-\omega_{ki} = \omega_{ik} = 
(\nu_in_k - \nu_kn_i)d\alpha\,$. Thus, in the vicinity of the identical transformation the matrices $\Lambda_k^i$ take the form $\Lambda_k^i(d\alpha)=\delta^i_k + {\omega^i}_k\,.$ Respectively, the $S(d\alpha)=1+\frac{1}{8} (\gamma^i \gamma^k - \gamma^k \gamma^i)\omega_{ik}\,$. Considering that $n_0=0$ and $\nu_0 = 1$, the latter relation leads to $S$ in the form:
\begin{equation}
\label{16}
S(\boldsymbol\nu;\boldsymbol n,\alpha) = e^{\left\{ \cdots \right\}{\alpha}/2}\,.
\end{equation}
Here and below, we use the notation $\left\{ \cdots \right\}$ for the sum of generators of Lorentz boosts and 3-rotations about the vector $[\boldsymbol\nu \boldsymbol n]\,$, that is 
\begin{equation}
\label{17}
\left\{ \cdots \right\} = -\,\gamma^0\boldsymbol\gamma\boldsymbol{n} - 
i\,\boldsymbol{\Sigma}[\boldsymbol{\nu} \boldsymbol{n}]\,,
\end{equation}
where $\,{\gamma}^0\,,\boldsymbol\gamma\,$ are the Dirac matrices, 
$\,\boldsymbol\Sigma =diag\,(\,\boldsymbol\sigma\,, \boldsymbol\sigma\,)\,$ 
and $\,\boldsymbol \sigma\,$ are the Pauli matrices. With the aid of the algebra of $\gamma$ matrices one can find that
\begin{equation}\nonumber
\begin{array}{lll}
\left\{ \cdots \right\}^2 = ({\boldsymbol \nu}{\boldsymbol n})^2 I\,, & \left\{ \cdots \right\}^4 =({\boldsymbol \nu}{\boldsymbol n})^4 I\,, & \cdots \,;\\
\left\{ \cdots \right\}^3 = ({\boldsymbol \nu}{\boldsymbol n})^3\left\{
  \cdots \right\}/({\boldsymbol \nu}{\boldsymbol n})\,, & 
\left\{ \cdots \right\}^5 = ({\boldsymbol \nu}{\boldsymbol n})^5\left\{
  \cdots \right\}/({\boldsymbol \nu}{\boldsymbol n})\,, & \cdots \,.\\ 
\end{array}
\end{equation}
These relations make it possible to represent (16) as 
\begin{equation}
\label{18}
S(\boldsymbol\nu;\boldsymbol n,\alpha) =I\, {\cosh} \frac{{\boldsymbol \nu}{\boldsymbol n}\alpha}{2} -
\, \frac{i\,\boldsymbol{\Sigma}[\boldsymbol{\nu} \boldsymbol{n}]+\gamma^0\boldsymbol\gamma\boldsymbol{n}}{{\boldsymbol \nu}{\boldsymbol n}}\, 
{\sinh}
\frac{{\boldsymbol \nu}{\boldsymbol n}\alpha}{2}\,.
\end{equation}
Now reparametrizing the $S(\boldsymbol \nu;\boldsymbol n,\alpha)$ with the aid of (8), (9) and using eqs.\ (15) and (13), we thus arrive at the following 3-parameter noncompact group of bispinor transformations in the axially symmetric flat Finslerian space (14): 
\begin{eqnarray}
\label{19}
\Psi'&=&\frac{\left ((1-\boldsymbol v\boldsymbol\nu )/\sqrt{1-{\boldsymbol v}^2}\right )^{-3r/2}}{2\sqrt{(1-\boldsymbol v\boldsymbol\nu )\sqrt{1-{\boldsymbol v}^2}}} \left\{\left (1-\boldsymbol v\boldsymbol\nu + \sqrt{1-{\boldsymbol v}^2}\right )I\right.  \nonumber \\
&&-\,\left. i\,[\boldsymbol{\nu} \boldsymbol{v}]\boldsymbol{\Sigma} - \left(\boldsymbol{v}-(1-\sqrt{1-{\boldsymbol v}^2})\,\boldsymbol\nu \right )\gamma^0\boldsymbol\gamma \,\right\}\, \Psi \,.
\end{eqnarray}
We note finally that an invariant of the transformations (19) is the Finslerian form 
$\,[({\nu_n \overline{\Psi}\gamma^n\Psi}/{\overline{\Psi}\Psi})^2]^{-3r/2}\overline{\Psi}\Psi\,$ but no longer the bilinear form $\overline \Psi \Psi$\,.

\bigskip
\bigskip

\noindent
{\large{\bf 4\,. Conclusion}}

\bigskip

\noindent
Having described the group of isometries of the axially symmetric
Finslerian event space (14) and the bispinor representation (19) of
its homogeneous 3-parameter subgroup (12), we have shown that in the case
of spontaneous breaking of the initial gauge symmetry an axially
symmetric fermion-antifermion condensate may arise. Such condensate
violates Minkowski geometry and, hence, Lorentz symmetry. However
relativistic symmetry remains valid. In the presence of condensate it
turns out to be represented by the 3-parameter group of generalized
Lorentz boosts.

In order to avoid misunderstanding it is important here to point out the
following. Relativistic symmetry is usually identified with Lorentz
symmetry, i.e. with symmetry under Lorentz boosts. At the same time it is
common knowledge that 3-parameter Lorentz boosts alone ( which belong to the 6-parameter Lorentz group ) do not make up a group. However, as was
shown in section 2, if each Lorentz boost is complemented by the
corresponding rotation of the space axes then the transformations thus
constructed already make up the 3-parameter subgroup (2) of the Lorentz
group. Linking different inertial reference frames, this subgroup includes
noncompact transformations only, and it is precisely this subgroup which,
strictly speaking, should be considered as a group of the relativistic
symmetry. From comparison of the group (2), acting in Minkowski space,
with the group (12), acting in the Finslerian space (14), it is seen that
these groups are locally isomorphic. Therefore the transformations (12),
referred to as generalized Lorentz transformations or generalized Lorentz
boosts, just realize a new representation of the relativistic symmetry
group.

 Since in all inertial reference frames, linked by the
generalized Lorentz boosts, the parameters of the axially symmetric
fermion-antifermion condensate remain invariant, any active
generalized Lorentz transformation of fundamental fields against the
background of such condensate is fully equivalent to the corresponding
passive transformation at which the condensate together with
fundamental fields is considered from another inertial frame. As a
result the generalized Lorentz symmetry is relativistic symmetry of
space-time filled with anisotropic fermion-antifermion condensate. In
this case its role is equally constructive as the role of usual
Lorentz symmetry in the standard theory of fundamental
interactions. In particular, the principle of generalized Lorentz
invariance makes it possible to take exactly into account the
influence of axially symmetric condensate on the dynamics of
fundamental fields. For example, the influence of the condensate 
on the dynamics of massive fermion field $\psi$ is specified by
the Lagrangian
\begin{equation}\nonumber
{\cal L}= \frac{i}{2}\left(\bar\psi{\gamma}^{\mu}{\partial}_{\mu}\psi -{\partial}_{\mu}
\bar\psi{\gamma}^{\mu}\psi\right) - m\!\left[\!\left(
\frac{\nu_{\mu} \bar{\psi} \gamma^{\mu} \psi}{\bar{\psi}\psi}
\right)^{\!\!\!2} \right]^{\!r/2}\!\!\bar{\psi}\psi  \,,
\end{equation}                                               
 where  $r$  and  $\nu_{\mu}=(1\,,\,-\boldsymbol\nu )$  are free relativistically invariant parameters
of the model. Being responsible for Lorentz symmetry violation,
these parameters characterize both anisotropy of Finslerian event
space (14) and properties of fermion-antifermion condensate, i.e.
the degree of its order and the direction $\,\boldsymbol\nu\,$ of its symmetry axis
respectively.
                                                                                                 
The above-written Lagrangian demonstrates the fact that the Finslerian
approach to the problem of Lorentz symmetry violation significantly
decreases (compared with SME) the number of free parameters which
describe possible effects of Lorentz symmetry violation. As for current
experimental bounds on Lorentz symmetry violation, in the literature
they are usually presented in terms of the SME Lorentz-violating
parameters (~see, for instance, [15]~). 
In connection with this a temptation may arise to consider the case where
the Finslerian parameter $\,r\,$ is equal to unity and then to identify the
$\,{\nu}_{\mu}\,$ with the parameters $\,a_{\mu}\,$ which appear, in particular,
in the neutrino sector [16] of the SME. Such identification however would
be incorrect since within the framework of the Finslerian model the case
$r=1$ corresponds to the unstable gauge symmetric state of the system of
interacting fields. At the same time it should be noted
that there exists a procedure which
allows us to express the Finslerian parameters $\,r\neq 1\,$ and $\,\boldsymbol\nu\,$ through the SME parameters
and thereby to obtain the corresponding experimental limits on the size
of Finslerian Lorentz-violating  parameters. However, such a procedure
requires cumbersome calculations. We hope to present them in our
subsequent publications.

\subsection*{Acknowledgements}

The author is grateful to Prof. H. Goenner for the fruitful collaboration that led 
to the results presented in this paper. Thanks are also due to him for a careful
reading of the manuscript and valuable remarks.

\end{document}